\documentclass[aip,reprint]{revtex4-1}
\usepackage{graphicx}
\usepackage{dcolumn}

\begin{document}


\title{Simulations of slow positron production using a low energy electron accelerator} 



\author{B. E. O'Rourke}
\email[]{brian-orourke@aist.go.jp}
\affiliation{National Institute of Advanced Industrial Science and Technology (AIST), AIST-Central 2, 1-1-1 Umezono, Tsukuba, Ibaraki 305-8568, Japan}

\author{N. Hayashizaki}
\affiliation{Tokyo Institute of Technology, Research Laboratory for Nuclear Reactors, 2-12-1 Ookayama, Meguro-ku, Tokyo 152-8550, Japan}

\author{A. Kinomura}
\author{R. Kuroda}
\affiliation{National Institute of Advanced Industrial Science and Technology (AIST), AIST-Central 2, 1-1-1 Umezono, Tsukuba, Ibaraki 305-8568, Japan}

\author{E. Minehara}
\affiliation{The Wakasa Wan Energy Research Centre, 64-52-1 Nagatani, Tsuruga, Fukui 941-0821, Japan}

\author{T. Ohdaira}
\author{N. Oshima}
\author{R. Suzuki}
\affiliation{National Institute of Advanced Industrial Science and Technology (AIST), AIST-Central 2, 1-1-1 Umezono, Tsukuba, Ibaraki 305-8568, Japan}



\begin{abstract}
Monte Carlo simulations of slow positron production via energetic electron interaction with a solid target have been performed. The aim of the simulations was to determine the expected slow positron beam intensity from a low energy, high current electron accelerator. By simulating (a) the fast positron production from a tantalum electron-positron converter and (b) the positron depth deposition profile in a tungsten moderator, the slow positron production probability per incident electron was estimated. Normalizing the calculated result to the measured slow positron yield at the present AIST LINAC the expected slow positron yield as a function of energy was determined. For an electron beam energy of 5 MeV (10 MeV) and current 240 $\mu$A (30 $\mu$A) production of a slow positron beam of intensity 5 $\times$ 10$^{6}$ s$^{-1}$ is predicted. The simulation also calculates the average energy deposited in the converter per electron, allowing an estimate of the beam heating at a given electron energy and current. For low energy, high-current operation the maximum obtainable positron beam intensity will be limited by this beam heating. 
\end{abstract}


\maketitle 

\section{Introduction}
The use of electron accelerators is a well established technique for the production of slow positrons, with numerous facilities based on linear accelerators (LINAC) \cite{Ley1997, Sueoka1985, Howell1987, Akahane1990, Tanaka1991, Ito1991, Kurihara2000, Chemerisov2007, Krause-Rehberg2008} or microtrons \cite{Mills1989, Merrison1999}. Positrons are generated via pair creation when energetic electrons are stopped in a high-Z target, the electron-positron converter. These positrons are then moderated, i.e. slowed to thermal energies, via interaction with a suitable material such as tungsten, and a slow positron beam produced. The positron production probability increases from threshold (1 MeV) with increasing electron energy and typically a value between 14 MeV and 100 MeV has been used.  

However, with the reduction in cost of low energy ($<$ 15 MeV), high-current industrial LINACs and development of new types of accelerator such as the Rhodotron\cite{rhodotron} it has become appealing to consider producing intense, slow positron beams using low energy, high current electron beams. One group, at Saclay, is developing such a system based on a 6 MeV commercial LINAC\cite{Perez2009, Muranaka2010}. 

\begin{figure}
\includegraphics[width=20pc]{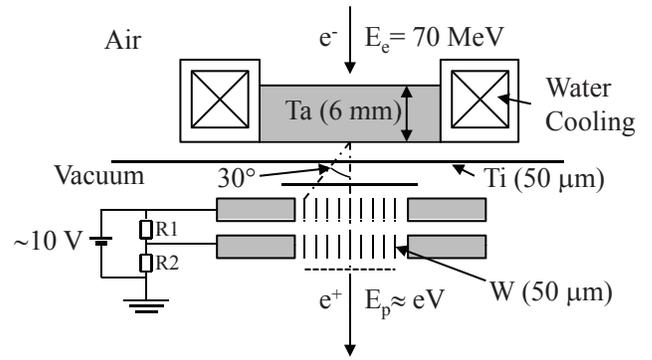}%
\caption{\label{AIST}A schematic diagram of the converter and moderator assembly currently in use at the AIST LINAC. Electrons at 70 MeV are directed on to a 6 mm thick, water cooled tantalum block in air. Positrons emerging in the forward direction pass through a thin titanium film into vacuum and are incident on an array of 50 $\mu$m tungsten films arranged in a rectangular mesh. Moderated positrons at low energy (several eV) are extracted from the moderator assembly by the applied positive potential ($\approx$10 V) and magnetically guided into a transport beamline.}%
\end{figure}

For the past 20 years the AIST LINAC has been used to produce high intensity, slow positron beams for materials research\cite{Akahane1990, Suzuki1997}. A schematic of the AIST electron-positron converter and moderator assembly is shown in figure \ref{AIST}. Electrons with a kinetic energy of 70 MeV are directed onto a water cooled tantalum disk with a thickness of 6 mm. Positrons emerging in the forward direction are incident on a moderator composed of strips of 50 $\mu$m tungsten foil arranged in a rectangular mesh. Slow (moderated) positrons are electrostatically extracted from the moderator then formed into a beam and magnetically guided to experimental stations located more than 20 m from (and well shielded from) the radiation produced in the converter.

The AIST LINAC operates at a frequency of 100 Hz, delivering a pulse of around 30 nC with a pulse width of $\approx 1 \mu$s. Slow positrons are produced with a similar time structure which is impractical for use in positron annihilation spectroscopy (PAS) type experiments such as Doppler broadening of annihilation radiation and positron annihilation lifetime spectroscopy (PALS)\cite{Coleman2000}. In practice, the beam is stored in a linear trap and transformed into a quasi-DC beam. In PALS this DC beam is chopped and bunched to form a pulse train with a pulse width of around 100 ps and a frequency of around 10 MHz.

This manipulation of the beam timing structure leads to a considerable loss of positrons, typically at most 10\% reach the sample in a PALS experiment. If the electron accelerator delivered smaller electron pulses with much higher frequency this situation could be improved. This is possible with a superconducting accelerator (SCA) and our group is now developing a dedicated SCA for positron production \cite{Hayashizaki2010, Orourke2010}. In an ideal case the positron beam produced in the converter/moderator could be used in PALS experiments without manipulation, resulting in an increased beam transport efficiency. 


However, in comparison to the beam energy used at the current AIST LINAC (70 MeV), the proposed SCA will be based on an accelerating module with a maximum acceleration of around 7.5 MV. Two such modules, previously used on the JAERI FEL project\cite{Kikuzawa1993, Minehara2006}, have been obtained by our group. Using these modules in a single pass configuration we expect a maximum beam energy of 7.5 MeV (1 module) or 15 MeV (2 modules). However, initially we expect to run with slightly lower than maximal field gradient with an expected beam energy of around 5 MeV per module. 

Using an accelerator with a low beam energy substantially reduces the positron production probability although it also leads to a much reduced radiation dose and induced activity from the accelerator. The maximum of the cross section for neutron production via the giant resonance occurs around 14 MeV\cite{Vylet2001} for tantalum. Keeping the electron energy below this maximum reduces the required radiation shielding dramatically, an advantage for compact, low-cost systems. 

The purpose of the present paper is to estimate the yield of slow positrons from the proposed SCA. We report results of Monte Carlo simulations of the converter and moderator using the Penelope2008\cite{Baro1995} code. Penelope2008 can perform calculations over a wide energy range, from a few hundred eV to about 1 GeV and provides full support for positron interactions. It is thus highly suited to modeling of positron production from energetic electron beams. 

Previously, several groups have reported Monte Carlo simulations of positron production from electron accelerators\cite{Mohri1991, Kossler1993, Segers1994, White1999, Perez2004}. In particular Gagliardi and Hunt\cite{Gagliardi2006} reported a systematic study of tungsten converters at normal and glancing irradiation for a range of incident electron energies. Most previous simulations have concentrated on the fast positron production from the converter. In recent years there has been progress in developing low energy extensions to the Monte Carlo code in order to fully simulate the moderation process\cite{Okada2000, Plokhoi2001}. In the present study both the converter and moderator are simulated using a standard code (without low energy extensions) by implementing a well established model of the moderation process.   

\section{Simulations}
\begin{figure}
\includegraphics[width=20pc]{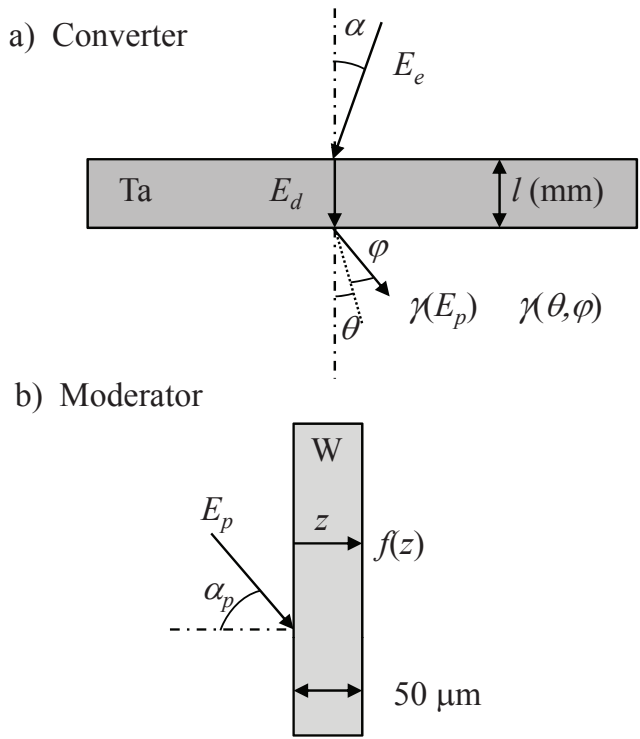}%
\caption{\label{target} Schematic of the (a) converter and (b) moderator simulation geometry with calculated output parameters in boldface. In (a) electrons of energy $E_e$, at an angle $\alpha$ to the normal, are incident on a cylindrical tantalum target with variable thickness, $l$, and the energy $\gamma(E_p)$, angular $\gamma(\theta, \phi)$ distributions of emerging positrons calculated. The average energy deposited in the converter per incident electron, $E_d$, is also calculated. In (b) mono-energetic positrons at an angle $\alpha_p$ are incident on a 50 $\mu$m tungsten foil and the positron depth deposition distribution, $f(z)$, determined. Using this distribution the probability of re-emission of a slow positron, $\eta$, is calculated as outlined in the text. By convolving the results of simulation (a) and (b), the probability of producing a slow (moderated) positron per incident electron, $\gamma_s$, can be estimated.}%
\end{figure}

A diagram of the simulation geometry is shown in figure \ref{target}. The simulation is divided into two parts which are simulated separately, (a) the converter and (b) the moderator. For each simulation between 10$^6$ and $10^8$ primary particle trajectories are calculated, with larger simulations typically necessary at lower electron energies when positron production probabilities are reduced. The low energy cut-offs were typically set at 50 keV for electrons and positrons and 5 keV for photons. Reducing the cut-offs below these values caused the computation time to increase without any significant change in the simulation results.

\subsection{The electron-positron convertor}
In this simulation an electron beam is incident on a cylindrical tantalum target. Input parameters are the electron energy, $E_e$, incident angle, $\alpha$, and convertor thickness, $l$. The output parameters of most interest are the angular, $\gamma(\theta,\phi)$, and energy, $\gamma(E_p)$ distributions of emerging positrons. Here $E_p$ is the energy of the emitted positron and $\theta, \phi$ the polar and azimuthal angles respectively. The polar angle is defined with respect to the axis perpendicular to the converter faces. The azimuthal angle is $\phi$ = 0 in the plane defined by the axis perpendicular to the converter face and the electron beam direction. When electrons are normally incident ($\alpha = 0$) the simulation is symmetric in the angle $\phi$. Finally we also note the average energy deposited in the converter per electron, $E_d$.



\subsubsection{Optimum converter thickness}
Initially the optimum converter thickness, $l_{opt}$, was determined. At each set of input parameters the probability (per electron) of producing a fast positron which leaves the converter in the forward direction ($0^\circ < \theta < 90^\circ$) was calculated, $\gamma_f$, and is given by,
\begin{equation}
\label{gamfegn}
 \gamma_f = \int^{2\pi}_0 \int^{\frac{\pi}{2}}_0 {\gamma(\theta,\phi)} {\text sin \theta} {\text d \theta} {\text d \phi}%
\end{equation}

\begin{figure}
\includegraphics[width=20pc]{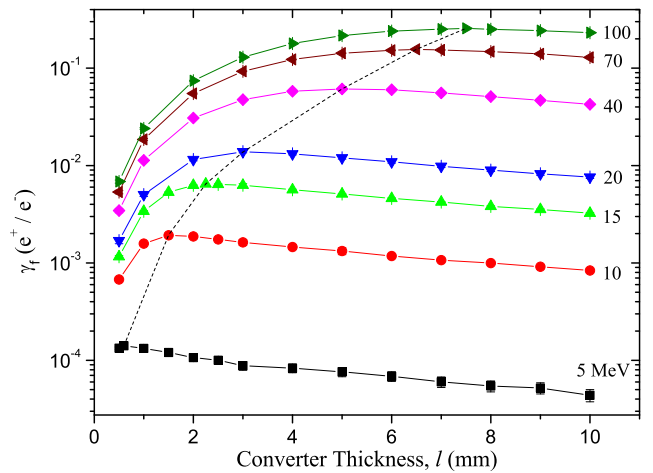}%
\caption{\label{tscan}(Color online) Variation in positron production probability in the forward direction, $\gamma_f$, with converter thickness, $l$, for a range of electron energies between $E_e$ = 5 MeV and 100 MeV. Electrons are normally incident on a tantalum converter in every case. The dashed line connects values at the optimal converter thickness, $l_{opt}$, at each electron energy.}%
\end{figure}

\begin{figure}
\includegraphics[width=20pc]{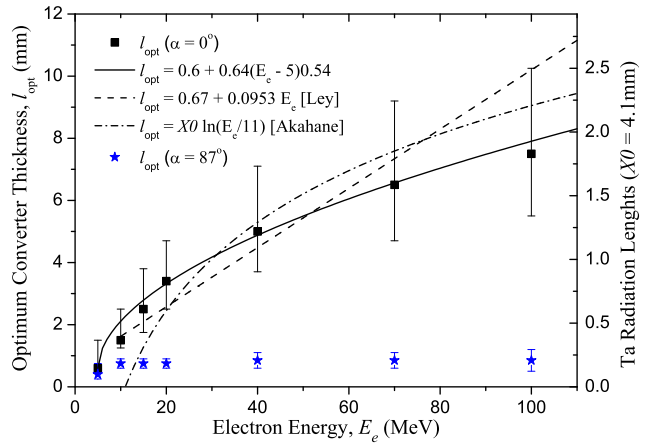}%
\caption{\label{tmax}(Color online) Optimum (maximum probability of fast positron production in the forward direction, $\gamma_f^{max}$) converter thickness, $l_{opt}$, calculated as a function of electron energy, $E_e$, at both normal ($\alpha = 0^\circ$) and glancing ($\alpha = 87^\circ$) incidence. The error bars represent the range of thickness over which $\gamma_f > 0.9 \gamma_f^{max}$. Also shown is an extrapolation based on previous measurements by Ley \cite{Ley2000} and the model of Akahane {\itshape et al.}\cite{Akahane1990}. }%
\end{figure}

Figure \ref{tscan} shows the results of a these simulations for a range of electron energies between $E_e$ = 5 MeV and 100 MeV and converter thickness up to 10 mm. At each electron energy the positron production probability, $\gamma_f$, increases to a maximum before decreasing with increasing converter thickness. This optimum thickness, $l_{opt}$, increases with increasing $E_e$ and the peak becomes broader. Figure \ref{tmax} shows a plot of the optimal thickness, $l_{opt}$ as a function of electron energy, $E_e$, for electrons at both normal ($\alpha = 0^\circ$) and glancing ($\alpha = 87^\circ$) incidence. The error bars represent not the statistical error in the calculation but rather the range over which $\gamma_f$ is $>90\%$ of the maximum.

At normal incidence, for the range of energies studied $l_{opt}$ varies from 0.6 mm at 5 MeV to 7.5 mm at 100 MeV. The calculated value at 70 MeV is 6.5 mm, close to the converter currently in use at AIST (6 mm). For glancing irradiation there is almost no variation in $l_{opt}$ with $E_e$, $l_{opt}$ changes from 0.4 mm at 5 MeV to 0.85 mm at 100 MeV. In all cases the value of $l_{opt}$ quoted is the real thickness of the converter material, i.e. not scaled by the angle of irradiation, $\alpha$. For normal incidence, a fit to these calculated data points allows us to estimate the optimum thickness, $l_{opt}$, of a Ta converter as a function of electron energy, $E_e$, as;
\begin{equation}
\label{optthick}
 l_{opt}[{\text{mm}}] = 0.6 + 0.64(E_e[{\text{MeV}}]-5)^{0.54}.%
\end{equation}

Also shown on figure \ref{tmax} is an extrapolation based on measurements of the optimum converter thickness at various laboratories\cite{Ley2000}, $l_{opt}$[mm] = 0.67 + 0.0953($E_e$[MeV]). This linear extrapolation tends to over-estimate the optimal thickness at high energy. Better agreement is seen with the simple model developed by Akahane {\itshape et al.} which assumes that the initial energy of the electron is shared between electron-positron and bremsstrahlung production\cite{Akahane1990}. They estimate an optimal thickness for any material based on the radiation length, $X0$, as;
\begin{equation}
 l_{opt} = X0 \frac{E_e}{E_c},%
\end{equation}  
where $E_c$ is the critical energy below which ionization becomes the dominant electron energy-loss mechanism. For tantalum $E_c$ = 11 MeV and $X0$ = 4.1 mm. This simple model provides a good estimate of $l_{opt}$ at high electron energy but underestimates the optimum value for $E_e <$ 20 MeV. However, for intermediate energies both the linear extrapolation and the Akahane model provide reasonable estimates of the optimal converter thickness. 
 
\subsubsection{Variation of fast positron production with electron energy}

\begin{figure}
\includegraphics[width=20pc]{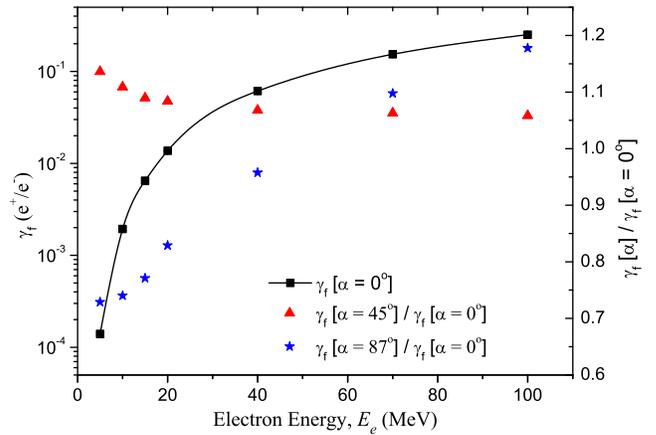}
\caption{\label{fastp}(Color online) Fast positron production probability in the forward direction, $\gamma_f$ at $l = l_{opt}$, calculated for several input electron energies $E_e$ for electrons normally incident ($\alpha = 0^\circ$) incident on the converter. Also shown in the relative increase or decrease in $\gamma_f$ for $\alpha = 45^\circ$ and $87^\circ$. Tantalum converters of optimal thickness were used for each simulation.}%
\end{figure}

Assuming optimal thickness converters are used we can then plot the positron production probability in the forward direction, $\gamma_f$, as a function of electron energy, $E_e$. Figure \ref{fastp} shows this plot for normally incident electrons ($\alpha = 0^\circ$) and shows a reduction in $\gamma_f$ by 3 orders of magnitude when $E_e$ is reduced from 70 to 5 MeV. Analysis for electrons incident at $\alpha = 45^\circ$ and $87^\circ$ is also shown on figure \ref{fastp}. The values are plotted as a comparison to the normally incident beam, i.e. $\gamma_f [\alpha]/\gamma_f [\alpha=0^\circ]$. In all cases the maximum value of $\gamma_f$ using converters of optimal thickness are plotted.  It is clear that while the optimal converter thickness $l_{opt}$ is sensitive to the angle of irradiation (figure \ref{tmax}), the maximum available fast positron intensity is less sensitive, provided optimized converters are used. At all energies $\gamma_f$ is increased when $\alpha = 45^\circ$ with a maximum increase of 13\% at for $E_e = 5$ MeV. Glancing irradiation ($\alpha = 87^\circ$) reduces $\gamma_f$ at low electron energies but actually leads to increased production for energies $> 70$ MeV. 

\subsubsection{Positron energy and angular distributions}

\begin{figure}
\includegraphics[width=20pc]{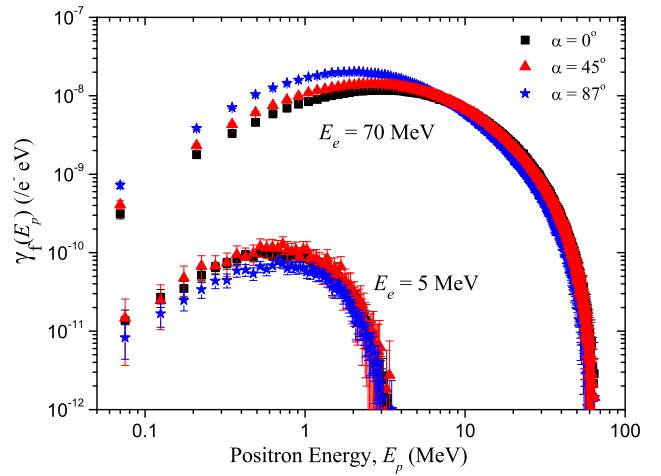}
\caption{\label{fastpdist}(Color online) Typical energy distributions of positrons emerging in the forward direction, $\gamma_f(E_p)$, after irradiation with 5 MeV and 70 MeV electrons at $\alpha$ = 0$^\circ$, 45$^\circ$ and 87$^\circ$. In all cases tantalum converters of optimum thickness, $l_{opt}$, were simulated.}%
\end{figure}

The calculated energy distribution of positrons emerging in the forward direction, $\gamma_f(E_p)$, is plotted in figure \ref{fastpdist} for $E_e$ = 5 MeV and 70 MeV for $\alpha$ = 0$^\circ$, 45$^\circ$ and 87$^\circ$. The energy distributions are very broad, covering the whole energy range from the low energy cut-off at 50 keV up to almost the initial electron energy $E_e$. For $E_e$ = 70 MeV the distribution is peaked around 2-3 MeV while at $E_e$ = 5 MeV the peak is around 0.6-0.7 MeV. 

\begin{figure}
\includegraphics[width=20pc]{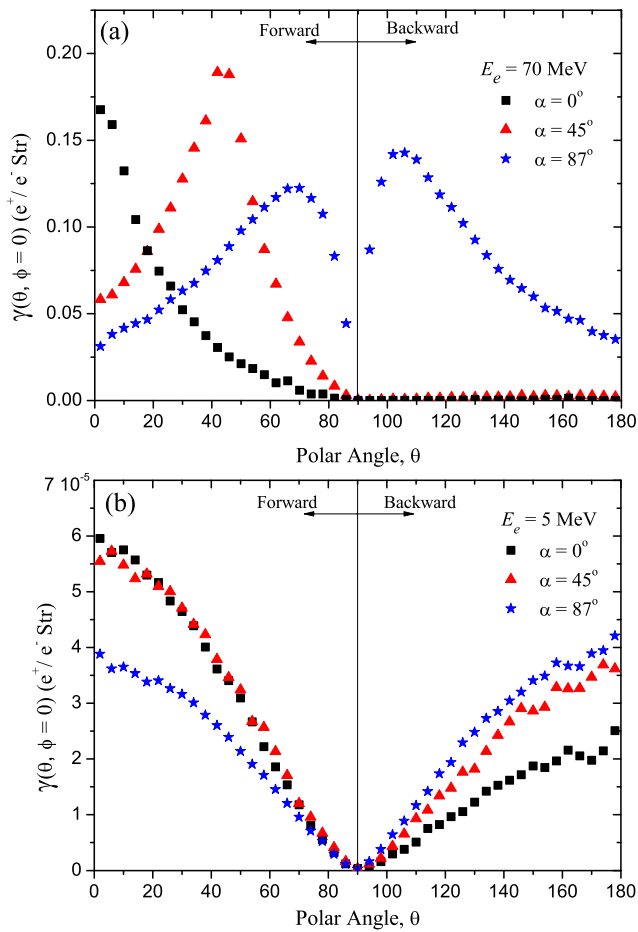}
\caption{\label{nonnormalangle}(Color online) Typical polar angular distributions in the central plane, $\gamma(\theta, \phi = 0)$, of emerging positrons after irradiation with (a) 70 MeV and (b) 5 MeV electrons at $\alpha$ = 0$^\circ$, 45$^\circ$ and 87$^\circ$. In all cases tantalum converters of optimum thickness, $l_{opt}$, were simulated.}%
\end{figure}

Figure \ref{nonnormalangle} plots the polar angular distribution of emerging positrons in the $\phi = 0$ plane after irradiation with (a) 70 MeV and (b) 5 MeV electrons at $\alpha$ = 0$^\circ$, 45$^\circ$ and 87$^\circ$. For non-normal irradiation the emerging positron distribution depends on both the polar ($\theta$) and azimuthal ($\phi$) angles. At $E_e$ = 70 MeV, the polar distribution is peaked in the same direction as the incident electron beam for $\alpha$ = 0$^\circ$ and $\alpha$ = 45$^\circ$. For $\alpha$ = 87$^\circ$ the peak is broader and has a maximum around $\theta$ = 70$^\circ$ and there is significant positron production in the backward direction, actually greater than the forward production in this case. 

For low energy irradiation ($E_e$ = 5 MeV) the polar distribution is much broader and peaked around the normal ($\theta = 0$) for all incident angles. Although not plotted, the azimuthal distribution is also more uniform at $E_e$ = 5 MeV compared to the case for $E_e$ = 70 MeV where the emerging positrons are emitted in a smaller cone around the axis. 

For normally incident electrons, $\alpha = 0$, the emerging positron angular distribution $\gamma(\theta, \phi)$ is symmetric about the azimuthal angle $\phi$ so $\gamma(\theta, \phi)$ can be described by the polar angle $\theta$ alone, $\gamma(\theta)$. For all energies the distributions are peaked on axis ($\theta = 0$) and have minima at $\theta = 90^\circ$. For $\theta > 90^\circ$ the positrons are emerging from the back of the converter in the opposite direction to the electron beam. This fraction is small at high energy but is comparable to the production in the forward direction for $E_e$ = 5 MeV. For high energy irradiation it was calculated previously that high energy positrons should be emitted with an angular distribution of the form\cite{Dahm1988} $\gamma_{\theta} = n_0 \exp(\theta/\theta_c)$, where $n_0$ is the intensity on axis ($\theta = 0$), and $\theta_c$ is a characteristic decay angle. This function was found to be a good description of $\gamma(\theta)$ at 100 MeV, but, for lower energies it was found to be necessary to introduce an offset in order to maintain an accurate fit, i.e. $\gamma(\theta) = n_0 [\exp(\theta/\theta_c) - A]$.%

\begin{figure}
\includegraphics[width=20pc]{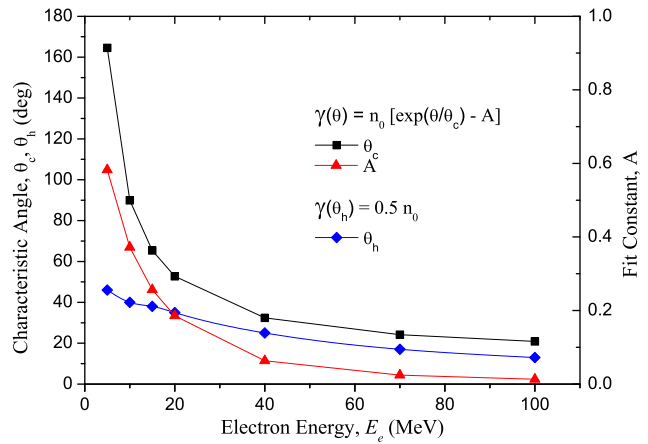}%
\caption{\label{angular}(Color online) Angular distribution of positrons in the forward direction from a Ta converter of optimal thickness when irradiated at normal incidence by electrons of energy $E_e$. The angular distributions were fitted with the function $\gamma(\theta) = n_0 [\exp(\theta/\theta_c) - A]$ and values of characteristic angle, $\theta_c$, and constant, $A$, are plotted. As the electron energy increases the positrons tend to emerge closer to the axis ($\theta = 0$). Also shown is the angle $\theta_h$ where the intensity falls to half that at $\theta = 0$, i.e. $\gamma(\theta_{h})$ = 0.5 $n_0$.}%
\end{figure}

Figure \ref{angular} shows the fitting parameters for the angular distribution as a function of $E_e$. As $E_e$ increases the offset tends to zero and $\theta_c$ decreases to around $20^\circ$. A second measure of the angular spread can be defined as the angle where the intensity drops to half that on axis, $\gamma(\theta_{h})$ = 0.5 $n_0$. This value is also plotted on figure \ref{angular} and shows a decrease from 46$^\circ$ at 5 MeV to 13$^\circ$ at 100 MeV. 


\subsection{The positron moderator}
The previous Monte Carlo simulations can help us estimate the total number and energy, angular distributions of positrons leaving the converter. In practice these positrons are then moderated, a process where the energetic positron is implanted into a material, thermalizes and may, with some probability diffuse to the surface and be re-emitted with a low (several eV) energy. At present we use an array of 50 $\mu$m tungsten films arranged in a rectangular mesh (figure \ref{AIST}).

Since it is impossible to simulate the full thermalization and diffusion process with the Monte Carlo code, a simple model is used as a basis for the present calculation of moderator efficiency.
Typically the moderation process is considered on the basis of a diffusion length, $L_+$, and re-emission branching ratio, $y_0$, for the moderator material. Since the probability that a positron can diffuse back to the surface from a given depth, $z$, is equal to $\exp(-z/L_+)$\cite{Charlton2001}, then the total probability of re-emission, $\eta$, is given by;
\begin{equation}
\label{modeq}
 \eta = y_0 \int^{\infty}_0{f(z) \exp(-z/L_+)} {\text dz}%
\end{equation}
where $f(z)$ is the positron deposition depth profile. 
  
\begin{figure}
\includegraphics[width=20pc]{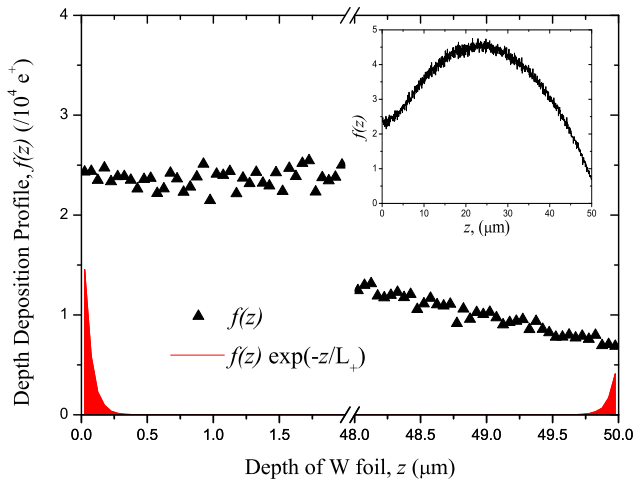}%
\caption{\label{5mev45}(Color online) Depth penetration profile, $f(z)$, of 0.5 MeV positrons incident at $\alpha_p = $45$^\circ$ on a 50 $\mu$m W foil. The solid line shows $f(z)$ scaled by the function $\exp{(-z/L_+)}$ which is equal to the probability of a positron diffusing back to the surface. The total area under this curve scaled by the re-emission branching ratio, $y_0$, determines the re-emission probability, $\eta$. The inset shows the full depth penetration profile, $f(z)$.}%
\end{figure}

For polycrystalline tungsten, Suzuki {\itshape et al.}\cite{Suzuki1998} measured a diffusion length, $L_+$ = 55 nm, and re-emission branching ratio, $y_0$ = 0.27. A schematic of the simulation geometry is shown in figure \ref{target}(b).  The implantation depth profile, $f(z)$, of positrons incident on W at a given energy, $E_p$, and angle, $\alpha_p$, is calculated by the simulation and values of $L_+$ and $y_0$ quoted above used to determine the re-emission probability, $\eta$, according to equation \ref{modeq}. Re-emission from both the front and back faces of the 50 $\mu$m foil was included in the analysis. An example of the calculated depth profile, $f(z)$, and the probability of diffusion to the W surface is shown in figure \ref{5mev45}. The re-emission probability is determined by calculating the total area under the solid curve, scaled by the re-emission branching ratio. 

\begin{figure}
\includegraphics[width=20pc]{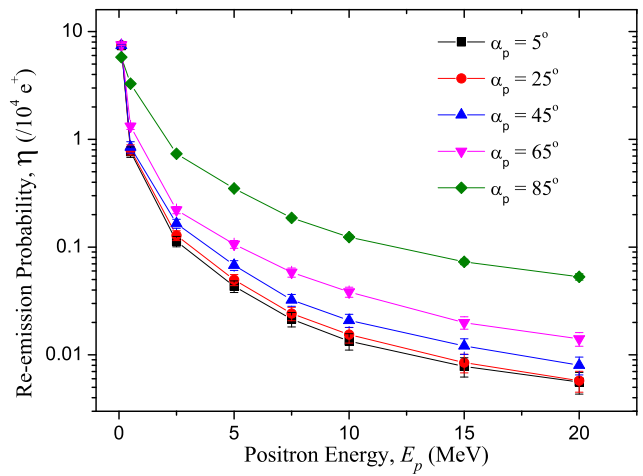}%
\caption{\label{modang}(Color online) Probability of positron re-emission, $\eta$, from a thin W foil (50 $\mu$m) as a function of incident positron energy, $E_p$, and angle of irradiation, $\alpha_p$.}%
\end{figure}

Results of this analysis are shown in figure \ref{modang} with the re-emission probability, $\eta$, plotted as a function of incident positron energy, $E_p$, for a range of incident angles, $\alpha_p$. It is clear that positrons incident on the moderator with low energy and at shallow angles have a much high re-emission probability than fast, normally incident particles.

Taking the angular and energy distributions of fast positrons from the previous converter simulation we can then estimate the slow (moderated) positron production probability, $\gamma_s$, at each initial electron energy. We consider only the case where electrons are normally incident on the converter. As a first step all positrons emitted at a polar angle $\theta >$ 30$^\circ$ from the normal are ignored, an assumption based on the geometry of the current converter and moderator. The ratio, R(30$^\circ$), is defined as the ratio of fast positrons emitted with angles less than or equal to 30$^\circ$ to the total forward (0-90$^\circ$) production, i.e. R(30$^\circ$) = $\int^{\pi/6}_{0} \gamma_f(\theta) {\text d \theta}/\int^{\pi/2}_{0} \gamma_f(\theta) {\text d \theta}$. R(30$^\circ$) decreases from 0.81 at 100 MeV to 0.57 at 5 MeV, for low energy operation it is therefore more important to locate the moderator as close as possible to the converter. 

We also assume that all fast positrons from the converter with $\theta <$ 30$^\circ$ are incident on a W moderator foil at an incident angle $\alpha_p$ = 90 - $\theta$ (since the converter and moderator faces are perpendicular, see figure \ref{AIST}). Finally it is assumed that all re-emitted positrons can be extracted to the slow positron beam. 
 
For the fast positrons with $\theta <$ 30$^\circ$, the fraction in each of the angular ranges, $\theta$ = 0-10$^\circ$, 10$^\circ$-20$^\circ$ and 20$^\circ$-30$^\circ$ was calculated and the energy distribution of this component (assumed to independent of angle) was convoluted with the moderation probability function at an angle of 85$^\circ$, 75$^\circ$ and 65$^\circ$ respectively and the contributions summed.

\begin{figure}
\includegraphics[width=20pc]{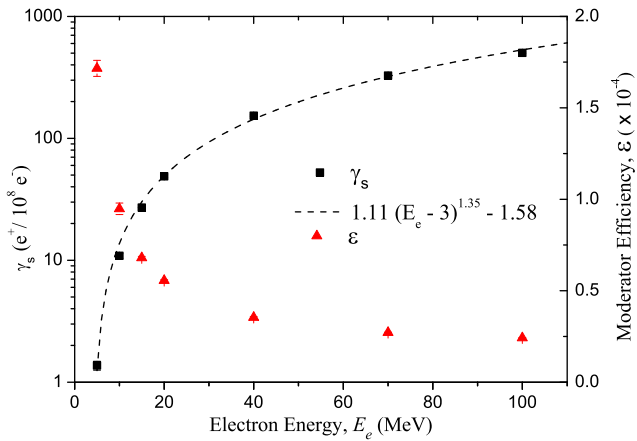}%
\caption{\label{slowp}(Color online) Probability of slow positron production per incident electron, $\gamma_s$, as a function of electron energy, $E_e$. The data points are based on the fast positron production probability, $\gamma_f$, calculated in the first simulation multiplied by a conversion factor which accounts for the moderation efficiency. Also shown is a fit to the calculations using a power law.}%
\end{figure}

The results of this conversion are plotted in figure \ref{slowp} which shows the slow (moderated) positron production probability, $\gamma_s$, as a function of $E_e$. Although the fast positron production probability is lower for lower energy incident electrons, the positrons created have a lower energy distribution (figure \ref{fastpdist}) and are hence moderated with higher efficiency than those created with a high energy beam. This result is somewhat negated by the broader angular distribution of the emitted fast positrons (figure \ref{nonnormalangle}), however on balance the efficiency of moderation tends to increase for lower energy irradiation. The moderator efficiency $\epsilon$ at a given energy is defined as;
\begin{equation}
  \epsilon = \frac{\gamma_s}{\gamma_f~R(30^\circ)}%
\end{equation}
and is also plotted on figure \ref{slowp}. $\epsilon$ decreases from 1.72 $\times$ 10$^{-4}$ at 5 MeV to 2.42 $\times$ 10$^{-5}$ at 100 MeV. 
 
The slow positron production probability, $\gamma_s$, at 5 MeV is around 0.4\% that at 70 MeV, an improvement of a factor of 4.6 compared to the same ratio for un-moderated, fast positrons due to the increased moderation efficiency, $\epsilon$. The calculated fast and slow positron production probabilities along with other relevant data are summarized in table \ref{tab1}.


\begin{table}
\caption{\label{tab1}Predicted fast, $\gamma_f$, and slow, $\gamma_s$, positron production probabilities from an optimal thickness tantalum converter and normal incidence electron irradiation. The ratio of fast positrons produced within 30$^\circ$ of the normal is also shown, R(30$^\circ$), along with the calculated moderation efficiency, $\epsilon$, based on the model described in the text. Numbers in round brackets indicate error estimates and square brackets order of magnitude.}
\begin{ruledtabular}
\begin{tabular}{cccccc}
E$_e$  & $l_{opt}$  & $\gamma _{f}$  & R(30$^\circ$) & $\gamma _{s}$ & $\epsilon$  \\
(MeV) & (mm) & (e$^+$/e$^-$) &   & (e$^+$/e$^-$) &  \\
\hline
5 & 0.6  & 1.41(04)[-4] & 0.57 & 1.37(12)[-8] & 1.72(04)[-4]  \\
10& 1.5  & 1.94(19)[-3] & 0.59 & 1.09(05)[-7] & 9.48(31)[-5] \\
15& 2.3  & 6.48(08)[-3] & 0.61 & 2.71(10)[-7] & 6.80(08)[-5] \\
20& 3.0  & 1.38(01)[-2] & 0.64 & 4.88(12)[-7] & 5.39(04)[-5] \\
40& 5.0  & 6.12(07)[-2] & 0.71 & 1.53(02)[-6] & 3.54(04)[-5] \\
70& 6.5  & 1.56(01)[-1] & 0.78 & 3.27(07)[-6] & 2.71(02)[-5] \\
100& 7.5 & 2.56(01)[-1] & 0.81 & 5.02(08)[-6] & 2.42(01)[-5] \\
\end{tabular}
\end{ruledtabular}
\end{table}

\section{Comparison to measured slow positron yields}

The maximum slow positron intensity from the present AIST facility was measured previously to be around 5 $\times$ 10$^7$ s$^{-1}$ at a beam energy of 70 MeV and current of 10 $\mu$A. This corresponds to a production probability of 8.35 x 10$^{-7}$ e$^+$/e$^-$. The simulation result at this energy is 3.27 $\times$ 10$^{-6}$, or around 4 times the measured value. There are several factors which account for this difference including; 1. Our assumption that all positrons emitted at a polar angle less than $30^\circ$ are incident on the moderator foil. Some fraction at very low angles may pass through without meeting any of the W foils; 2. We also assume that every positron which is moderated can be extracted to the slow positron beamline. Clearly some fraction will not be guided out of the moderator assembly due to re-collisions with the W surface; 3. Degradation of the moderator due to electron induced defects and surface contamination.  

\begin{figure}
\includegraphics[width=20pc]{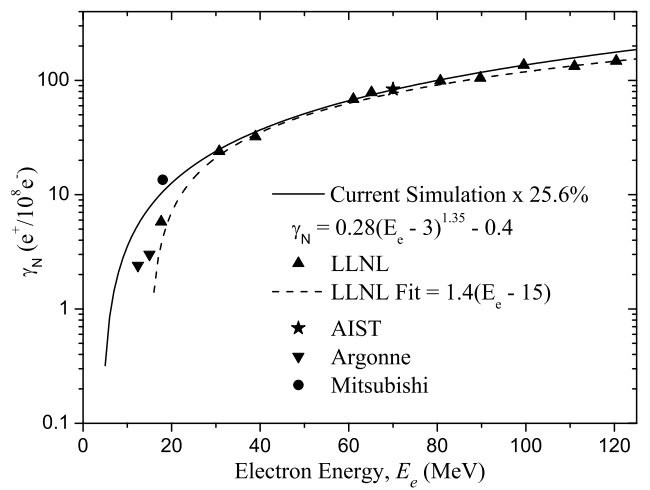}%
\caption{\label{normalized}Simulation result normalized to the AIST measurement at 70 MeV plotted alongside previous measurements at LLNL\cite{Howell1987} and Argonne\cite{Chemerisov2007} and Mitsubishi Electric Corporation\cite{Tanaka1991}. The LLNL and Argonne measurements tend to underestimate the positron production at low energy compared to the present simulation however much better agreement is seen with the 18 MeV measurement at Mitsubishi.}%
\end{figure}

In order to compare the current simulation results to the measured value at 70 MeV the simulation result was normalized by scaling by a factor of 25.6\%. This normalized result, $\gamma_N$, is plotted in figure \ref{normalized} along with several previously reported measurements\cite{Howell1987, Tanaka1991, Chemerisov2007}.
	
The normalized simulation result shows good agreement with the measured results at LLNL especially at high energy. The LLNL LINAC used a Ta convertor and W moderator assembly similar to that considered here. The measurements had a minimum energy of 17.5 MeV and were taken with a fixed assembly, i.e. fixed convertor thickness of 5 mm, rather than an optimized convertor. The authors assumed a simple linear electron energy relationship with a cut-off at 15 MeV, $\gamma_s [e^+/ 10^8 e^-] = 1.4 (E_e{\text{[MeV]}} - 15)$,(fit from\cite{Okada1990}).


The current simulation suggests a much higher production probability at lower energy ($E_e <$ 20 MeV) than that predicted by the linear LLNL fit. This disagreement is also apparent in the low energy Argonne measurements\cite{Chemerisov2007, Chemerisov2007b}. Much closer agreement is seen with the measurement at the Mitsubishi Electric Corporation LINAC\cite{Tanaka1991} where a 5 mm Ta converter and W moderator were used and an estimated slow positron production efficiency of 1.35 $\times 10^{-7}$ s$^{-1}$ at 18 MeV was reported. 

\section{Beam heating of the converter}

\begin{figure}
\includegraphics[width=20pc]{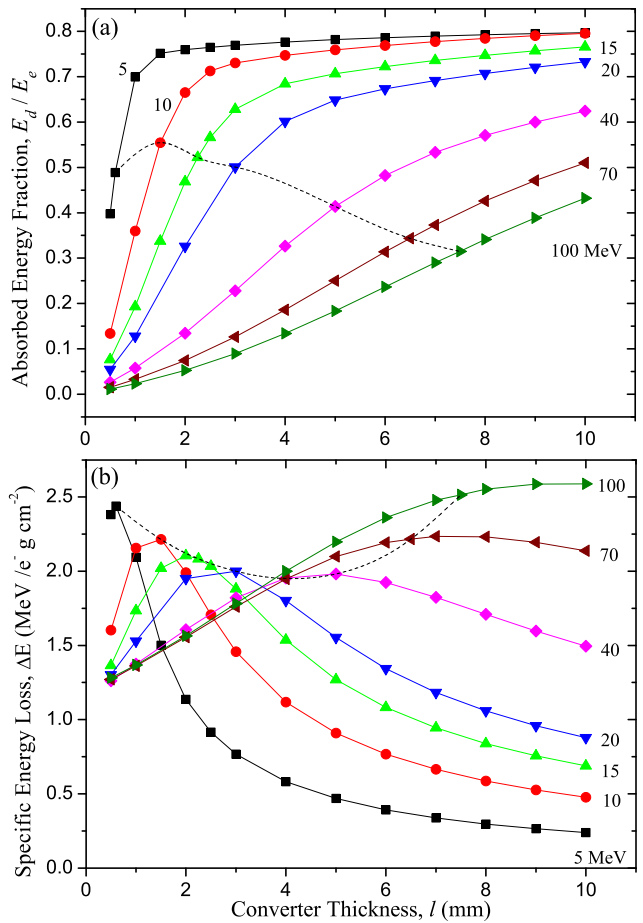}
\caption{\label{energyloss}(Color online) Plots of (a) the fractional energy absorption in the converter, $E_d/E_e$, and (b) the specific energy loss in the converter per incident electron, $\Delta E$, for an electron beam at normal incidence impinging on a tantalum converter as a function of thickness, $l$. The dashed lines connect values at the optimal converter thickness, $l_{opt}$, at each electron energy.}%
\end{figure}

Based on the previous calculations it is then straightforward to estimate the expected slow positron intensity at a given electron beam energy and current. It is also clear that if intense slow positron beams are to be generated using a low energy accelerator then a high beam current is required. However, at this stage we need to consider the energy deposited in the converter.  The ratio of average energy deposited in the converter per electron, $E_d$, divided by the electron energy $E_e$ is plotted in figure \ref{energyloss}(a) as a function of the converter thickness, $l$, for a normally incident electron beam. This ratio determines the fraction of the beam power deposited in the converter and increases with increasing converter thickness up to a maximum. For the range of energies studied this maximum is around 0.8, showing that even with very thick converters around 1/5 of the input power is not deposited in the converter but instead escapes in the form of backscattered and transmitted electrons, positrons, and high energy photons.  

The energy deposition in the converter can also be described by the specific energy loss parameter, $\Delta E$, which is defined as,
\begin{equation}
 \Delta E = \frac{E_d}{l~\rho} [{\text{MeV/e~g~cm}}^{-2}],%
\end{equation}  
where $l$ is the converter thickness and $\rho$ is the material density. $\Delta E$ is plotted in figure \ref{energyloss}(b). At low electron energy $\Delta E$ quickly reaches a maximum and subsequently decreases rapidly with increasing thickness. As the electron energy is increased the thickness at which this maximum is reached increases and is found to be similar the thickness at which positron production in the forward direction reaches a maximum, $l_{max}$. At low energy the maximum value of $\Delta E$ is greater than 2 MeV/e$^-$~g~cm$^{-2}$, decreasing to around 2 MeV/e$^-$~g~cm$^{-2}$ at intermediate energy, before increasing again for $E_e \geq$ 70 MeV. 

As the electron beam current increases the energy deposited will eventually be sufficient to melt or vaporize the converter material. An estimate for the cooling power of a water cooled converter, $P_w$, is given by the following equation \cite{Andreani1975},
\begin{equation}
\label{temp}
 P_w =  \frac{(T_b - T_w)2\pi\lambda l}{(0.5 + \ln(r_2/r_1))}
\end{equation}
where $r_2$ and $r_1$ are the radius of the converter and electron beam respectively, $\lambda$ is the thermal conductivity and $l$ the converter thickness. $T_b$ is the maximum temperature of the material inside the electron beam (on axis), while $T_w$ is the temperature of the copper block in contact with the water coolant. For tantalum $\rho$ = 16.7 g/cm$^3$ and $\lambda$ = 57.6 W/(m~K) at 300 K. The thermal conductivity of tantalum has been measured to increase with increasing temperature, up to a value of 61.4 W/(m~K) at 1800 K\cite{Savchenko2008} but this small temperature dependence was ignored in the subsequent analysis and the room temperature value used.    

The converter material will also radiate heat, with the power dissipation, $P_r$, given by,
\begin{equation}
\label{rad}
 P_{r} = 2 e \sigma(\int_{0}^{r_2}{(T(r)^4 - T_0^4)  2 \pi r {\text dr})} 
\end{equation}
where $T_0$ is the temperature of the surrounding environment and $\sigma$ is the Stefan-Boltzmann constant (5.67 $\times 10^{-8}$ J s$^{-1}$ m$^{-2}$ K$^{-4}$). $e$ is the spectral emissivity which depends on the condition of the materials surface, for un-polished tantalum it was measured to be around 0.32 over the temperature range 1100 - 2300 K\cite{Milosevic1999}. $T(r)$ is the radial temperature profile and is given by, 
\begin{eqnarray}
T(r) &=& T_b - \frac{P}{4 \pi \lambda l}\frac{r^2}{r_1^2}~~\{0 < r < r_1\}\nonumber \\
     &=& T_b - \frac{P}{2 \pi \lambda l}[0.5 + {\text{ln}}(r/r_2)]~~\{r_1<r<r_2\}
\end{eqnarray}

At equilibrium the power deposited in the converter by the electron beam is equal to the sum of the power dissipated by the water cooling and radiative losses. 
\begin{widetext}
\begin{equation}
\label{temprise}
 E_d I_e = P_w + P_{r}  =  \frac{(T_b - T_w)2\pi\lambda l}{(0.5 + \ln(r_2/r_1))} + 2 e \sigma (\int_{0}^{r_2}{(T(r)^4 - T_0^4)  2 \pi r {\text dr}})  
\end{equation}
The maximum electron current $I_{max}$ is given by,
\begin{equation}
\label{imax}
 I_{max} = \frac{C(T_m - T_w) +  D[\int_{r_0}^{r_2}{(T(r)^4 - T_0^4) }  2 \pi r {\text dr}]}{E_d} %
\end{equation}
\end{widetext}
where $T_m$ is the melting point of tantalum (3270 K) and $C$ and $D$ are constant equals to $\frac{2\pi\lambda l}{(0.5 + \ln(r_2/r_1))}$ and $2 e \sigma $ respectively.

It is clear that if the electron accelerator can provide sufficient electron current the maximum number of positrons at a particular electron energy, $Y_{max}$, is equal to the product of the slow positron production probability, $\gamma_s$, and the maximum possible electron current, $I_{max}$, i.e.,
\begin{equation}
\label{fom}
  Y_{max} = \gamma_s~I_{max} \propto \frac{\gamma_s}{E_d} \propto \frac{l~\gamma_s}{\Delta E}.%
\end{equation}
In practice the yield will be lower than $Y_{max}$ as it is unpractical to operate at $I_{max}$. The effective maximum current will be some fraction of $I_{max}$.

\section{Expected beam intensity}

\subsection{Optimal thickness converters, $l_{opt}$}
\begin{figure}
\includegraphics[width=20pc]{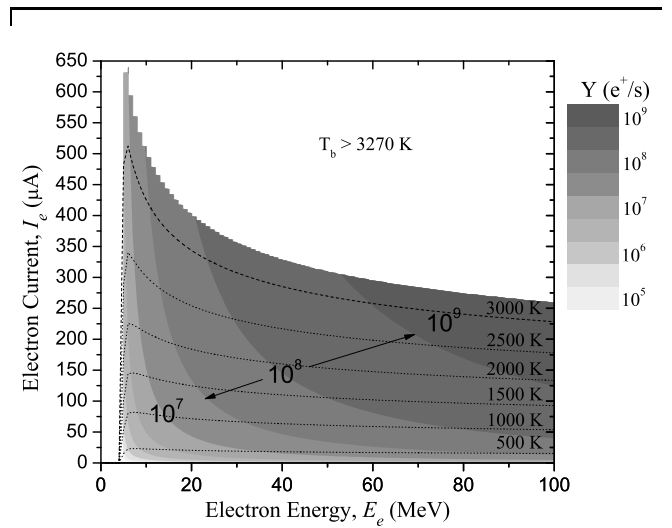}%
\caption{\label{2dplot}Surface plot showing the calculated slow positron yield, ($Y$), as a function of incident electron energy $E_e$ (x-axis) and current $I_e$ (y-axis). Ta converters of optimal thickness $l_{opt}$ and the normalized slow positron production probability $\gamma_N$ (figure \ref{normalized}) are used to calculate $Y$. The temperature of the converter material inside the electron beam $T_b$ was also calculated according to equation \ref{temprise} and is also shown. The high-current cut-off is determined as the current at which the Ta  converter melts ($T_b$ $>$ 3270 K)}%
\end{figure}

Figure \ref{2dplot} shows a 2D surface plot of the expected slow positron yield, $Y$, as a function of electron beam energy and current according to, $Y = \gamma_N~I$ . At each set of input parameters the temperature of the converter material inside the electron beam $T_b$ was also estimated using equation \ref{temprise} with $T_w = 350$ K, $T_0 = 300$ K, $r_1 = 1$ cm and $r_2 = 2$ cm. The thickness of the converter was also varied with energy according to equation \ref{optthick}. 

A comparison of the present AIST LINAC and possible future operation with a higher current lower energy SCA is shown in table \ref{tab2}. The present accelerator operates at 70 MeV and a maximum current of 10 $\mu$A. The table also shows the results for a high current beam (200 $\mu$A). With a 5 MeV electron beam a positron yield of 4.2 x 10$^6$ s$^{-1}$ is expected. A 200 $\mu$A (beam power = 1 kW) beam is well within the capabilities of the proposed SCA, therefore the main limitation to increasing the positron yield is the power deposited in the Ta convertor. The simulations suggest that at these conditions the power deposited is around 488 W. Using equation \ref{temprise} for a  water cooled converter with the dimensions given above ($r_1 = 1$ cm and $r_2 = 2$ cm) the estimated maximum temperature the converter material inside the electron beam, $T_b$, is around 2280 K. It is clear that effective cooling of the converter is essential at these high currents. 

\begin{table}
\caption{\label{tab2}Calculated values of the normalized slow positron production probability, $\gamma_N$, and energy deposited in a tantalum converter, $E_d$, for normal incidence electron irradiation on converters of optimal thickness, $l_{opt}$. Values of the of the slow positron beam intensity, $Y$, and temperature, $T_b$, of the converter at an electron current of 10 $\mu$A and 200 $\mu$A are also included. Numbers in square brackets indicate order of magnitude.}
\begin{ruledtabular}
\begin{tabular}{ccccccc}
    &      &       &  \multicolumn{2}{c}{10 $\mu$A}  & \multicolumn{2}{c}{200 $\mu$A} \\
E$_e$  & $\gamma _{N}$  & $ E_d $   & Y & $T_b$  & Y & $T_b$ \\
(MeV)  & (e$^+$/10$^8$e$^-$) & (MeV/e$^-$) & (e$^+$/s)  & (K) & (e$^+$/s) & (K) \\
\hline
5 & 0.35   & 2.44 & 2.10[5] & 434 & 4.19[6] & 2277 \\
10& 2.78   & 5.55 & 1.66[6] & 389 & 3.33[7] & 2296 \\
15& 6.91   & 7.83 & 4.14[6] & 394 & 8.28[7] & 2234 \\ 
20& 12.5   & 10.0 & 7.47[6] & 399 & 1.49[8] &  2239 \\
40& 39.2   & 16.6 & 2.35[7] & 411 & 4.70[8] &  2297 \\
70& 83.5   & 24.1 & 5.00[7] & 423 & 1.00[9] &  2525 \\
100& 128.6 & 31.5 & 7.68[7] & 431 & 1.50[9] &  2790 \\
\end{tabular}
\end{ruledtabular}
\end{table}

\subsection{Variation with converter thickness}
\begin{figure}
\includegraphics[width=20pc]{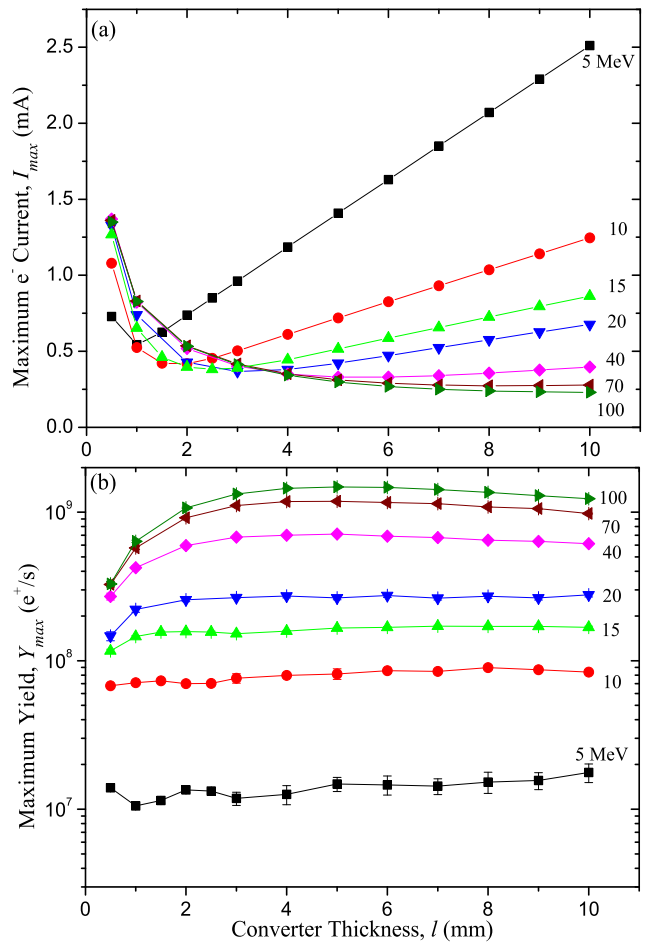}
\caption{\label{ymax}(Color online) (a) Maximum electron current, $I_{max}$, based on equation \ref{temp} and (b) maximum positron yield, $Y_{max}$, based on the product of the $\gamma_s$ and $I_{max}$ as a function of converter thickness, $l$, for electrons normally incident on a tantalum converter.}%
\end{figure}

Figure \ref{ymax} shows plots of (a) the maximum current $I_{max}$ and (b) the maximum yield as a function of converter thickness for a range of electron energies between 5 and 100 MeV. The simulations suggest that if the electron current from the accelerator is not a limiting factor then increased positron production is possible if thicker than optimal converters are used at low energies ($E_e <$ 20 MeV).  At lower energy, increasing the thickness beyond optimal has little impact on the fractional energy absorption (already near saturation), hence the maximum possible electron current increases and negates the slight reduction in $\gamma_s$. Conversely, for high electron energies the optimum converter thickness for positron production, $l_{opt}$, occurs at a point on the fractional energy absorption curve (figure \ref{energyloss}(a)) below the saturation level. This means that increasing the thickness beyond optimal values also increases the energy deposition and the hence lowers the maximum possible electron current.    

\section{Summary}
In conclusion, the present Monte Carlo simulations of the convertor and moderator provide an estimate of the slow positron yield as a function of electron energy and current provided by an electron accelerator in the range 5 $< E_e <$ 100 MeV. Although there is an exponential decrease in the positron production probability from the electron-positron converter as the electron energy is reduced, the lower energy distribution of the emitted positrons results in slightly higher moderation efficiencies. Intense slow positron beams can be generated if high electron currents are used, for example at 10 MeV and 200 $\mu$A an intensity of 3.3 $\times$ 10$^7$ e$^+$/s is predicted, similar to the typical intensities achieved with existing high energy, low current facilities.

The limiting factor for high current operation may not be the accelerator performance but rather the power deposited in the converter. For low energy, high current accelerators, the maximum slow positron yield may be increased by using converters of greater than optimal thickness. The design of the converter and moderator assembly for any low energy, high current accelerator will need to consider the beam heating of both the converter and moderator, rotating targets will improve the maximum possible beam power but their design and implementation would add further complexity to the system.

\begin{acknowledgments}
This work was supported by JSPS KAKENHI 21340087.
\end{acknowledgments}

%
\end{document}